\documentclass[sort&compress,3p]{elsarticle}
\usepackage{amsmath}
\usepackage{amsfonts}
\usepackage{amssymb}
\usepackage{graphicx}
\usepackage{subfig}
\usepackage{bbm}
\pdfoutput=1

\journal{Annals of Physics}

\begin{document}

\begin{frontmatter}

\title{Quantum phase transition by employing trace distance along with the density matrix renormalization group}

\author[ad1,ad2]{Da-Wei Luo}

\author[ad1]{Jing-Bo Xu\corref{cor}}
\ead{xujb@zju.edu.cn}
\cortext[cor]{Corresponding author}

\address[ad1]{Zhejiang Institute of Modern Physics and Department of Physics, Zhejiang University, Hangzhou 310027, People's Republic of China}
\address[ad2]{Beijing Computational Science Research Center, Beijing 100084, People's Republic of China}

\begin{abstract}
We use an alternative method to investigate the quantum criticality at zero and finite temperature using trace distance along with the density matrix renormalization group. It is shown that the average correlation measured by the trace distance between the system block and environment block in a DMRG sweep is able to detect the critical points of quantum phase transitions at finite temperature. As illustrative examples, we study a spin-1 XXZ chains with uniaxial single-ion-type anisotropy and the Heisenberg spin chain with staggered coupling and external magnetic field. It is found that the trace distance shows discontinuity at the critical points of quantum phase transition and can be used as an indicator of QPTs.
\end{abstract}


\end{frontmatter}

\section{Introduction}

Quantum phase transition (QPT) has attracted much attention in condensed matter physics~\cite{Sachdev1999} and become a hot topic over the years. Traditional QPT approaches mainly focus on the identification of the order parameters and the pattern of symmetry breaking. The existence of a QPT strongly influences the behavior of many-body system near the critical point associated with the divergence of correlation length of two-point correlation functions and the vanishing of the gap in the excitation spectrum. QPTs are a qualitative change in the ground state properties of a quantum many-body system as some external parameters of the Hamiltonian is varied. However, absolute zero temperature is unattainable experimentally due to the third law of thermodynamics. Therefore, it is of great theoretical and experimental importance to detect a QPT at finite temperatures~\cite{Werlang2010} low enough where thermal fluctuations are small enough so that quantum fluctuations still dominate. Recently, the trace distance~\cite{Nielsen2000,Gilchrist2005} approach has been developed with some success in the study of the quantum phase transition in a coupled cavity lattice at finite temperature~\cite{Luo2013}. This approach which compares a state and its factorized state, is different from the fidelity approach~\cite{Gu2010}, which compares two ground states whose Hamiltonian parameters are slightly varied, and allows us to detect QPTs without any prior knowledge of order parameters or the pattern of the symmetry breaking.

As more realistic or more complex systems are taken into consideration, analytic methods may not always work or severe approximations have to be made. Therefore, numerical tools can be a very valuable supplement when studying such systems. Over the years, many different tools have been proposed to study many-body systems such as the cluster Monte-Carlo methods~\cite{Kandel1991}, exact diagonalization~\cite{Bonner1964} and especially the density matrix renormalization group (DMRG)~\cite{Schollwock2011,Schollwock2005,White1992} which is generally regarded as one of  the most powerful numerical method available in the study of one-dimensional quantum lattice systems. The DMRG method is an iterative algorithm to optimally obtain an effective Hamiltonian along with all operators and wave functions of interest in a truncated set of basis states so that all relevant physics of the targeted state can still be captured. In this paper we present a method to study the QPT at finite temperatures using the average correlation between the system and environment block measured by the trace distance in conjunction with the DMRG method. Targeting the lowest-lying states, we can obtain an approximation of the Gibbs state at low temperatures. At every step of the DMRG sweep, we calculate the Gibbs state $\rho$ of the super-block along with the marginals of the system and environment block $\rho_{S(E)}=Tr_{E(S)}[\rho]$. The average correlation is obtained in one complete DMRG sweep by calculating the average trace distance between the Gibbs state of the super block $\rho$ and the tensor product of the marginals $\rho_S\otimes\rho_E$. Our approach may be a better candidate than the fidelity approach for use with DMRG because the set of basis states obtained by DMRG is dependent on the parameter of the Hamiltonian and generally it's not so easy to find the relationship between the two set of basis states which means that the inner product of two states cannot be directly calculated. This problem is non-existent in our approach as the states compared are obtained in a single DMRG sweep. As an illustrative example, we first investigate a spin-1 XXZ chains with uniaxial single-ion-type anisotropy at zero temperature. At each step of the sweep procedure of DMRG, we calculate the correlation between the system block and the environment block using the trace distance, and the result agree with the phase diagram obtained suing exact diagonalization. In order to show how to use the trace distance along with the density matrix renormalization group to detect quantum phase transitions at finite temperature, we study a Heisenberg spin chain with both staggered coupling and staggered magnetic field as an example. At the critical points of the quantum phase transition, which are determined by the ground state energy level crossing, the trace distance shows a sudden change of value at finite temperature, indicating that the trace distance can be used to describe the critical points of QPT and increasing the anisotropic parameter causes the phase boundary to shift towards smaller external magnetic fields. In order to confirm the validity of our approach, the geometric phase of the ground state is also calculated and it agrees well with the results obtained using our trace distance approach.
\begin{figure}
\begin{centering}
  \includegraphics[scale=.6]{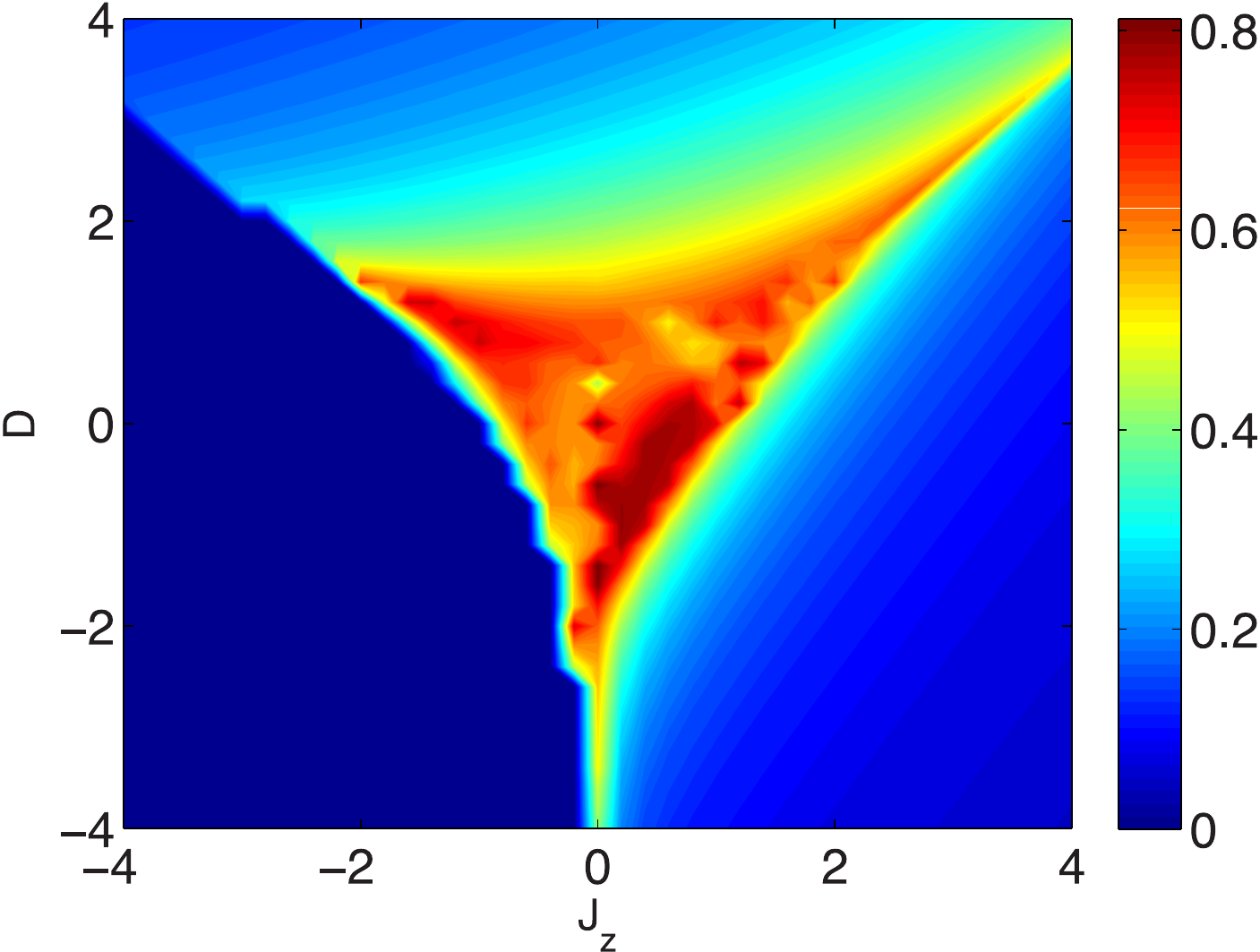}\\
  \caption{(Color online) Average trace distance for a spin-1 chain of length $N=150$ as a function of uniaxial single-ion anisotropy $D$ and coupling strength $J_z$. It can be observed that at the critical points of QPT, the trace distance shows a sudden change in value.}\label{fig_s1}
  \end{centering}
\end{figure}

\begin{figure*}
\begin{centering}
  \includegraphics[scale=.9]{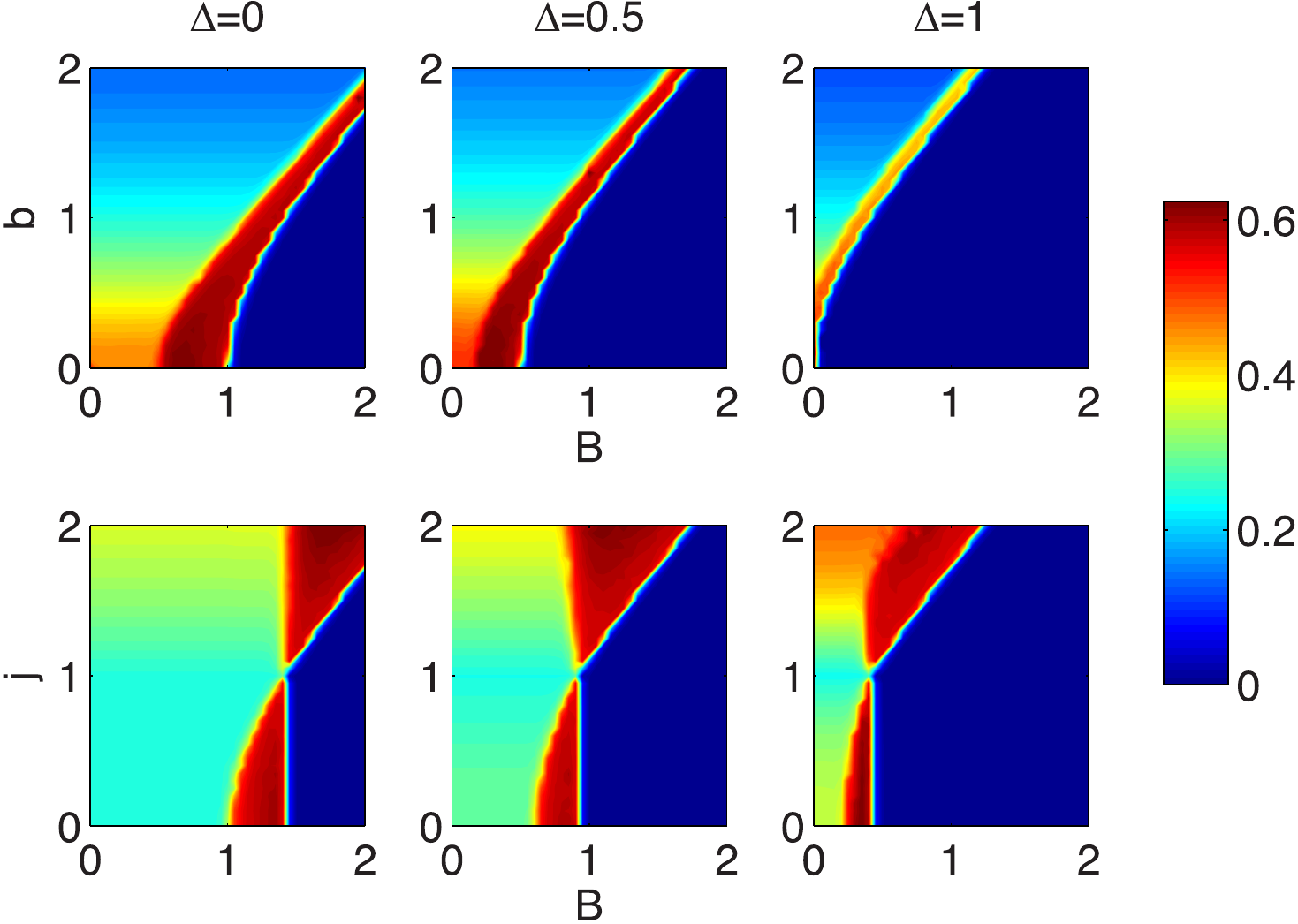}\\
  \caption{(Color online) Average trace distance for varying anisotropic parameter $\Delta$ at finite temperature $\beta=40$ as functions of magnetic field $B$ and alternating field parameter $b$ with $j=0.5$ (upper) and average trace distance as functions of magnetic field $B$ and alternating coupling parameter $j$ with $b=1$ (lower) with $N=150$. It can be observed that at the critical points of QPT, both show a sudden change in value and the phase boundaries shift to smaller fields as $\Delta$ gets bigger.}\label{f1}
  \end{centering}
\end{figure*}


\section{Trace distance approach to study QPT}\label{sec_td}
The trace distance between the density operator of a composite system and the Kronecker product of its marginals has recently been shown to be able to witness initial system-environment correlations in open-system dynamics~\cite{Smirne2010,Laine2010} and to distinguish quantum states and been used to detect QPT in a cavity QED array~\cite{Luo2013}. The distance of two trace class operators $\rho_1$ and $\rho_2$ is defined to be half the trace norm of $\rho_1-\rho_2$. For density operators, the trace distance can be further simplified as
\begin{equation}
D(\rho_1,\rho_2)=\frac{1}{2}\sum_i|d_i|\label{dst_def0},
\end{equation}
where $d_i$ are the eigenvalues of $\rho_1-\rho_2$. The trace distance ranges from zero to one, with its being zero if and only if the two states are identical. It is also a metric on the space of physical states and is sub-additive with respect to the tensor product,
\begin{equation*}
D(\rho_a\otimes\rho_1,\rho_b\otimes\rho_2)\leq D(\rho_a,\rho_b)+D(\rho_1,\rho_2).
\end{equation*}
We present a method to employ the trace distance measure in conjunction with the density matrix renormalization group (DMRG) to detect the critical point of quantum phase transition. We first break the composite system into a system block $S$ and an environment block $E$, which is also the first step in a typical DMRG calculation. The ground state or Gibbs thermal state can be described by a density matrix $\rho_{SE}$, which corresponds to the super-block in the DMRG framework. Next, we calculate the reduced density matrix of the system and the environment block, denoted $\rho_S$ and $\rho_E$ respectively. The reduced density matrix reads
\begin{align*}
\rho_S&=\mathrm{Tr}_E[\rho_{SE}]=\sum_{f_i}\langle e_i|\rho_{SE}|e_i\rangle,\\
\rho_E&=\mathrm{Tr}_S[\rho_{SE}]=\sum_{a_i}\langle s_i|\rho_{SE}|s_i\rangle,
\end{align*}
where $\mathrm{Tr}_E$ and $\mathrm{Tr}_S$ means the partial trace and is carried out by tracing over all environmental (system) basis $|e_i\rangle$ ($|s_i\rangle$) to obtain the marginal for the system (environmental) block. Define $\widetilde{\rho}=\rho_S\otimes\rho_E$, and it is obvious that $\rho_{SE}$ and $\widetilde{\rho}$ have the same marginals for the system and environment, $\mathrm{Tr}_{E(S)}[\rho_{SE}] =\mathrm{Tr}_{E(S)}[\widetilde{\rho}]$, then the difference between the two density matrices measured by the trace distance $D(\rho,\widetilde{\rho})$ can capture the system-environment correlation between the generic thermal Gibbs state of the super-block and the product of the marginals. This approach fits very well in a typical finite system DMRG sweep calculation since the reduced density matrix for the system or environment block is already calculated. To improve numerical accuracy, we take the average value of the trace distance at every step in a complete sweep procedure where the system or environment block size grows from $2$ to $L-2$ and shrinks back to $2$. As the system undergoes a quantum phase transition, the structure of the ground state or the thermal Gibbs state is significantly changed, and we expect to find a sudden change in value for the average correlation measured by the trace distance.

The fidelity approach~\cite{Gu2010} has been very successfully used to detect quantum criticality in a wide range of models. For a system described by a Hamiltonian of the from
\begin{equation*}
	H=H_0+\lambda H_I,
\end{equation*}
where $H_I$ is the driving Hamiltonian and $\lambda$ denotes its strength, the fidelity between the ground states $|\psi(\lambda) \rangle$ and $|\psi(\lambda+\delta\lambda)\rangle$ defined by their inner product has been found to show a sudden drop at the critical points of QPT, where $\delta\lambda$ is a small displacement. However, this approach may not be the best fit for use with DMRG due to the fact that the DMRG method is in effect an algorithm to find a set of basis states so that the targeted state, generally the lowest-lying eigenstates of the Hamiltonian, can be approximately expressed in a greatly reduced Hilbert space. Therefore, this set of basis states is generally dependent on the parameter of the Hamiltonian such as the strength of the external field applied. As a result, the set of basis  states obtained using DMRG for $H(\lambda)$ may not be the same as $H(\lambda+\delta\lambda)$. The standard approach of using fidelity with DMRG~\cite{Langari2013,Tzeng2012} is, therefore, one needs to obtain the ground state of each of the two involved Hamiltonians with slightly different parameters in the same DMRG step, and use both ground states as target states for updating the basis. With our approach, we eliminates this need for target states since only the reduced density matrix for the system and environment block is needed for any given set of external parameters, which are already calculated in one DMRG sweep. Another advantage of our method over the fidelity is that we do not require a small change in the external parameter $\delta\lambda$ and could have better numerical stability.

\begin{figure}
\begin{centering}
  \includegraphics[scale=.9]{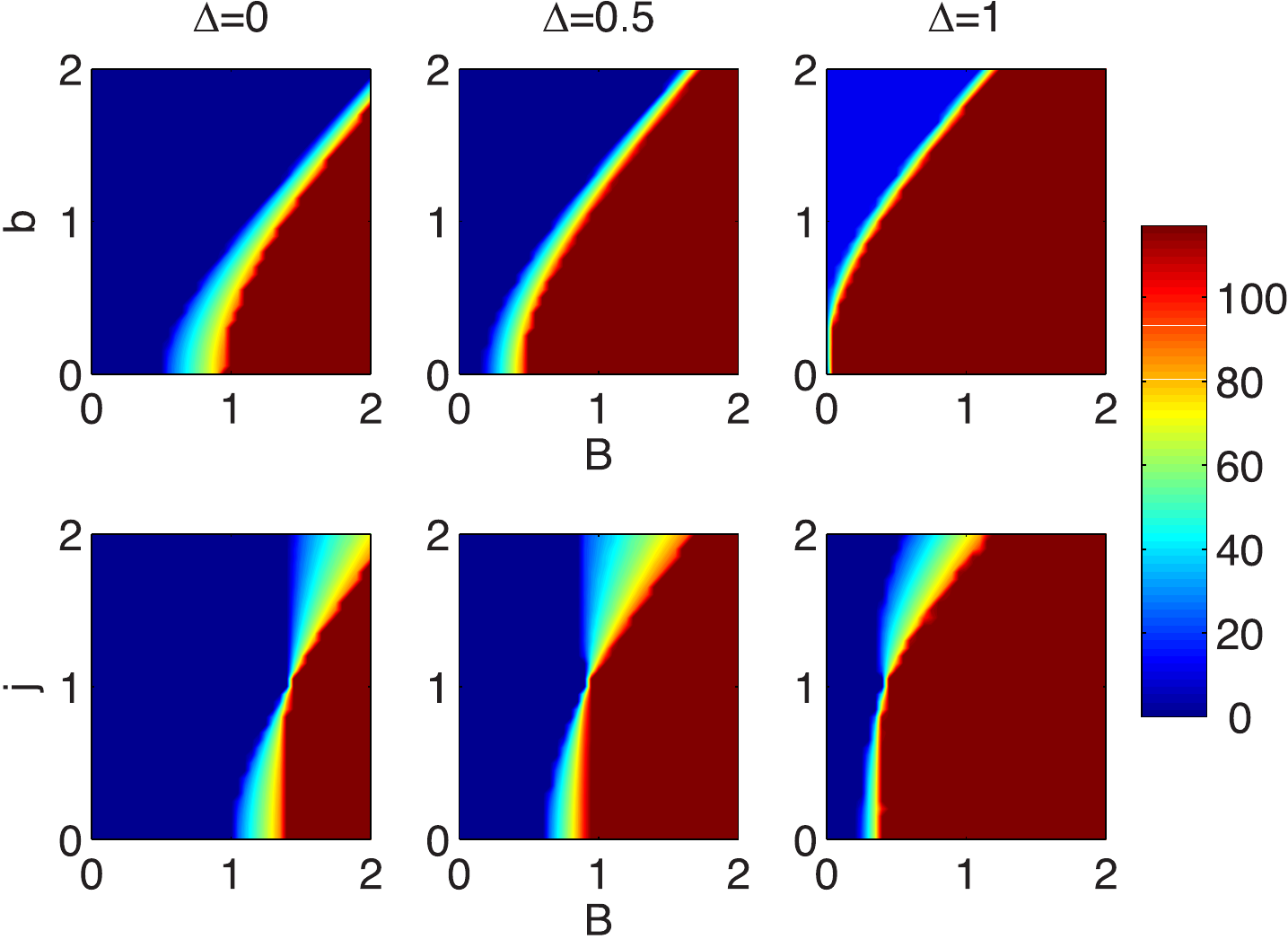}\\
  \caption{(Color online) Geometric phase for varying anisotropic parameter $\Delta$ at finite temperature $\beta=40$ as functions of magnetic field $B$ and alternating field parameter $b$ with $j=0.5$ (upper) and geometric phase as functions of magnetic field $B$ and alternating coupling parameter $j$ with $b=1$ (lower) with $N=150$. It can be observed that at the critical points of QPT, both show a sudden change in value and the phase boundaries shift to smaller fields as $\Delta$ gets bigger.}\label{f2}
  \end{centering}
\end{figure}

\section{Spin-1 XXZ chains with uniaxial single-ion-type anisotropy}\label{sec_ex_s1}

As an illustrative example, we investigate the quantum criticality in a spin-1 XXZ chains with uniaxial single-ion-type anisotropy by employing the trace distance along with the density matrix renormalization group in this section. Spin chains have been the subject of numerous investigations. The uniform anti-ferromagnetic Heisenberg spin chain is known to have a gapless ground state for half integer spins such as spin-1/2 chains~\cite{Bethe1931}. However, there exists a gap for integer spins called the Haldane gap~\cite{Affleck1989}. This state is affected by various perturbations such as single-ion-type anisotropy~\cite{Chen2003,Langari2013,Liu2013,Manmana2011}. The Hamiltonian of the spin-1 XXZ chains with uniaxial single-ion-type anisotropy is given by
\begin{equation*}
	H=\sum_i (S_i^x S_{i+1}^x+S_i^y S_{i+1}^y+J_z S_i^z S_{i+1}^z)+D\sum_iS_i^z{}^2,
\end{equation*}
where $J_z$ is the relative coupling along the z-axis, $D$ is the uniaxial anisotropic parameter and $S$ are spin-1 operators. This spin chain model has attracted a lot of interest because it has a very rich phase diagram including the Haldane phase, the XY phase, the ferromagnetic phase and the Neel phase~\cite{Chen2003,Schulz1986,Nijs1989}. An open boundary condition is also assumed. At zero temperature, we can find out the ground state wave function of the Hamiltonian using DMRG. We calculate the trace distance between the density operator of the ground state wave function $\rho=|\psi_G\rangle\langle \psi_G|$ and the direct product of the system-environment blocks 
\begin{eqnarray*}
	\widetilde{\rho}&=&\rho_S\otimes\rho_E,\\
	\rho_{S(E)}&=&\mathrm{Tr}_{E(S)}[\rho],
\end{eqnarray*}
during a complete sweep and obtain its average value within the DMRG sweep. Because the trace distance defined above can serve as an average measure of quantum correlation between the system and environment block, we expect it to show singular behavior at the critical points of QPT due to an abrupt change in the structure of the ground state wave function, enabling us to see the phase boundaries of QPT. We plot the average trace distance as a function of uniaxial single-ion anisotropy $D$ and coupling strength $J_z$ in Fig.~\ref{fig_s1} with $14$ target states. It can be readily seen that at the average trace distance shows a discontinuity at the critical points of QPT, in good agreement with the phase diagram obtained using numerical diagonalization and other DMRG studies~\cite{Chen2003,Langari2013,Liu2013,Manmana2011}.

\section{Spin-1/2 chain with staggered field and coupling at finite temperature}\label{sec_ex}

\begin{figure}[t]
\begin{centering}
  \includegraphics[scale=.56]{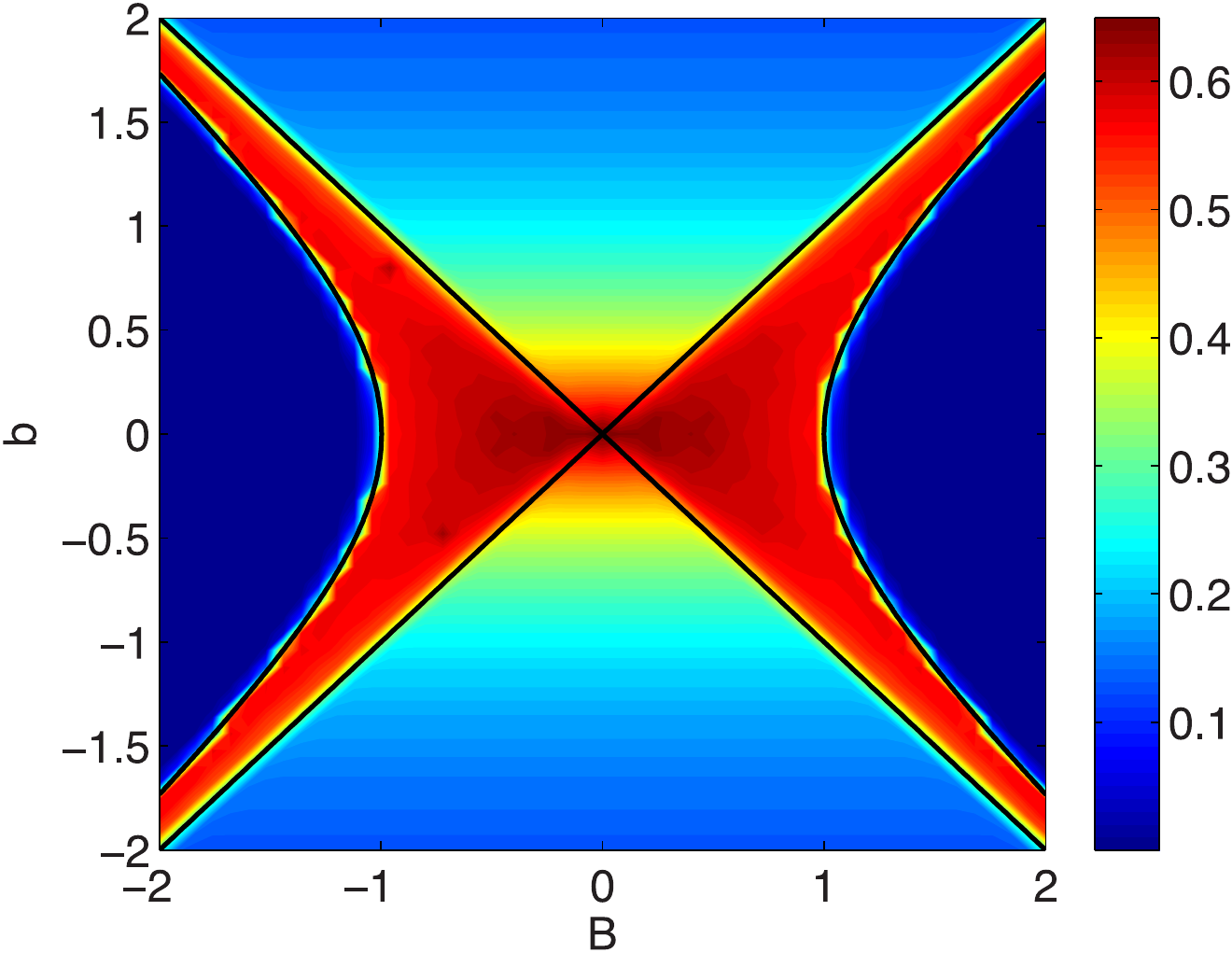}
  \caption{(Color online) Trace distance as a function magnetic field $B$ and alternating field parameter $b$ with $\Delta=0$ and chain length $N=150$ at finite temperature $\beta=40$. The solid black lines are the phase boundaries obtained from analytical results. It can be observed that at the critical points of QPT, the trace distance shows a sudden change in value, in good agreement with analytic results.}\label{f3}
\end{centering}
\end{figure}

In this section, we apply our method on a Heisenberg spin chain with staggered field and coupling\cite{Hide2012,Hide2007,Huang2013,Wang2013} as an example to illustrate its quantum criticality at finite temperature. The system under consideration is described by the Hamiltonian
\begin{equation*}
	H=-\sum_{l=1}^{N}\left[\frac{J_l}{2}\left(\sigma_l^x \sigma_{l+1}^x+\sigma_l^y \sigma_{l+1}^y+\Delta\sigma_l^z \sigma_{l+1}^z\right)+B_l \sigma_l^z\right],
\end{equation*}
where $\sigma$ are the Pauli matrices, $\Delta$ is the anisotropic parameter , $J_l=J+e^{i\pi l}j$ is the staggered coupling between spins and $B_l=B+e^{i\pi l}b$ is the staggered external magnetic field. It is noted that the spin chain system includes a wide variety of interesting spin chain systems such as the non-ergodic scaling behavior studied in~\cite{Deng2008}. In  the special case of $\Delta=0$ the system becomes the staggered $XY$ spin chain model which is exactly solvable~\cite{Hide2012}, and we can compare our results with the analytical results. Because QPTs are generally only detectable at very low temperatures, we can obtain an approximation of the Gibbs state $e^{-\beta H}/Z$ by targeting the lowest-lying states with DMRG, where the inverse temperature $\beta=1/k_b T$, $k_b$ the Boltzmann constant and $Z=\mathrm{Tr}(e^{-\beta H})$ is the partition function. Following our approach outlined above, we calculate the correlation between the system block and the environment block using trace distance $D(\rho,\rho_S\otimes\rho_E)$ during every renormalization step in a finite-size DMRG sweep procedure, and take its average value. At higher temperatures, the thermal fluctuation will mask the quantum fluctuation, making the detection of QPTs more difficult. As a result, we mainly focus on very low temperatures. By comparing the result obtained from 14 target states with the analytic exact result when the the anisotropic parameter $\Delta=0$ for the spin 1/2 chain with staggered field, we conclude that such choice can give a good approximation of the phase boundary at a relatively low computation cost. We have calculated the trace distance using finite-size DMRG up to $5$ sweeps after which there is little change between sweeps and the result converges. The maximum discarded weight is on the order of $10^{-4}$. In the upper panels of Fig.~\ref{f1} we plot the average trace distance at finite temperature with the inverse temperature $\beta=40$ and $j=0.5$ for varying anisotropic parameter $\Delta$ as functions of magnetic field $B$ and alternating field parameter $b$ with chain length $N=150$ with $14$ target states. It can be seen that at the critical points the average trace distance shows a sudden change in value, which means that our method can be used to detect the critical points of QPTs. The trace distance for varying anisotropic parameter $\Delta$ at finite temperature as functions of magnetic field $B$ and alternating coupling parameter $j$ is shown in the lower panels of Fig.~\ref{f1}, and it can also be observed that at the critical points of QPT, a sudden change in value is observed and the phase boundaries shift to smaller fields as $\Delta$ gets bigger. In both cases, when $\Delta=0$, our results agree exactly with the analytical results in Ref.~\cite{Hide2012,Hide2007}, and the phase boundaries shift to smaller fields as $\Delta$ gets bigger. From Fig.~\ref{f1} , it can also be observed that below the critical points of QPT, the quantum correlation measured by the trace distance stays constant as the static external magnetic field $B$ varies, showing that the static magnetic field $B$ is not a dominant parameter below the critical point. As QPTs generally occur due to the competition between the strength of the coupling constants and that of the magnetic fields, the dominate parameter for the staggered Heisenberg model is the coupling parameter below the critical points and the magnetic parameter above the critical point. On the other hand, the correlation is sensitive to the alternating magnetic field parameter $b$ below the critical points, illustrating the difference between the two magnetic field parameters $B$ and $b$.

At the critical points of QPTs, the structure of the ground state is drastically altered, changing the geometry of the Hilbert space. Therefore, the geometric phase as a measure of the curvature of the Hilbert space should be able to capture this behavior and display critical behavior~\cite{Carollo2005}. As a comparison, we also calculate the geometric phase in the special case of $T\rightarrow 0$ obtained by the rotation~\cite{Carollo2005}
\begin{equation}
	H(\phi)=g(\phi)Hg^\dagger(\phi) \: \mathrm{with} \: g(\phi)=\prod_{l=1}^{N}\exp[i \sigma_l^z \phi/2].\label{hrot}
\end{equation}
The geometric phase $\varphi_g$ obtained under the rotation given Eq.\eqref{hrot} is given by
\begin{equation*}
\varphi_g=-i\int_0^{\pi}\langle g|\frac{\partial}{\partial \phi}|g\rangle
\end{equation*}
We plot the geometric phases for both cases in Fig.~\ref{f2}, and it can be readily shown that the phase boundary obtained using the average trace distance approach agree extremely well with the geometric phase, and our approach is more easily obtained using DMRG at finite temperatures. It's also worth noting that as we allow both the coupling strength and the external magnetic field to be alternating and the interaction along the $z$ axis is also taken into consideration through the anisotropic parameter $\Delta$, our model includes a wide variety of interesting spin chain systems such as the non-ergodic scaling behavior studied in~\cite{Deng2008}. We plot the average trace distance as a function of the magnetic field strength $B$ and the field alternating parameter $b$ with the critical points obtained analytically superimposed in Fig.~\ref{f3}. It can be readily seen that along the critical points of QPT, the average trace distance in the special case of $\Delta=0$ can serve as an indicator of quantum criticality as it shows a drastic change in value, in good agreement with the analytic predictions. It is interesting to study finite size scaling behavior of the trace distance or its derivative and compare it with other signatures of quantum phase transitions such as fidelity.

\section{Summary}\label{sec_sum}

In conclusion, we have applied the trace distance along with the density matrix renormalization group to explore the quantum criticality at zero and finite temperature. It is shown that by using the trace distance between the system block and environment block in a DMRG sweep as a measure of their average correlation, we are able to pinpoint the critical points of quantum phase transitions at finite temperature. As an illustrative example, we first study a spin-1 XXZ chains with uniaxial single-ion-type anisotropy. It is found that the trace distance displays a discontinuity at the critical points of QPT and its phase diagram is also obtained. We then also investigate the spin chain model with staggered coupling and external magnetic field at finite temperature. It is observed that both the average trace distance and the geometric phase show discontinuity at the critical points of quantum phase transition at finite temperature. It is worth pointing out that our model includes a wide variety of interesting spin chain systems and in the special case of $\Delta=0$, our results agree very well with the analytic results obtainable for the isotropic staggered XY spin chain. The method developed in this present paper can be employed to study a variety of many-body systems and may help us gain a deep understanding of QPTs at finite temperature. In particular, the matrix product states (MPS) based DMRG algorithms~\cite{Schollwock2011} are well suited to calculate finite temperature properties, and it would be very interesting to apply the idea in this paper to the MPS framework and study finite temperature QPTs in various systems. 

\section{Acknowledgment}

This project was supported by the National Natural Science Foundation of China (Grant No. 11274274).

\bibliographystyle{elsarticle-num}


\end{document}